# Effects of disease duration and antipsychotics on brain age in schizophrenia

Alejandro Roig-Herrero[a,b,*], Luis M. San-José-Revuelta[b,c], Rafael Navarro-González[b,d], Rodrigo de Luis-García[b], Vicente Molina[a,e]

[a]Psychiatry Department, School of Medicine, University of Valladolid, Valladolid, Spain
[b]Imaging Processing Laboratory, University of Valladolid, Valladolid, Spain
[c]Castilla-León Neuroscience Institute, University of Salamanca, Salamanca, Spain
[d]Neurology Service, Clinical Hospital of Valladolid, Valladolid, Spain
[e]Psychiatry Service, Clinical Hospital of Valladolid, Valladolid, Spain

## Abstract:

Accelerated brain aging has been consistently reported in patients with schizophrenia. Over the past decade, these findings have been replicated using the Brain Age paradigm, which applies machine learning techniques to estimate brain age from neuroimaging data. This approach yields a single index, the Brain Age Gap, defined as the difference between predicted and chronological age. Nevertheless, both the progressive nature of this phenomenon and the potential role of antipsychotic medication remain unclear.

To investigate its progression, we compared the Brain Age Gap between individuals experiencing a first episode of psychosis and healthy controls using ANCOVA, adjusting for age, sex, body mass index, and estimated total intracranial volume. To enhance the robustness of our findings, we employed two distinct models: a transformer-inspired model based on harmonized volumetric brain features extracted with FastSurfer, and a previously trained deep learning model. To assess the potential effect of medication, we further compared bipolar patients who received antipsychotic treatment with those who did not. Mann-Whitney U test consistently showed that medicated bipolar patients did not exhibit a significantly larger Brain Age Gap.

Both models converge on the conclusion that accelerated brain aging is unlikely to be explained by antipsychotic medication alone. Longitudinal studies are therefore required to clarify the temporal dynamics of brain aging in schizophrenia.

Abbreviations: Magnetic Resonance Imaging (MRI), Body Mass Index (BMI), First Episodes (FE), estimated Total Intracranial Volume (eTIV), mean absolute error (MAE)

# 1. Introduction

Patients with schizophrenia consistently exhibit structural brain alterations, including diminished grey matter in the frontal and temporal lobes as well as in the hippocampus, as along with an enlargement of the third ventricle (Brugger and Howes, 2017; Haijma et al., 2013; Shepherd et al., 2012). In this context, it has been proposed that patients with schizophrenia experience accelerated aging or exhibit alterations that closely resemble those associated with early aging. This phenomenon has been substantiated through biomarkers such as oxidative stress, retinal degeneration, gene expression or synaptic function (Nguyen et al., 2018; Seeman, 2022; Shivakumar et al., 2014). This apparent accelerated aging is also observed in the brain (Blose et al., 2023; Chiapponi et al., 2013) and has been linked to cognitive decline in mood disorders (Ho et al., 2024).

Among various methodologies, the brain age paradigm (Franke et al., 2010) has emerged as a promising approach for providing an age-adjusted measure of structural brain health, which is especially relevant for investigating the intersection of aging and schizophrenia-related changes. This paradigm has been utilized in schizophrenia and other psychiatric and non-psychiatric conditions to examine the link between aging and disease (Ballester et al., 2022; Lieslehto et al., 2021). In this approach, a machine learning model is trained on neuroimaging data from healthy individuals to predict their chronological age. The deviation between an individual's actual age and the predicted age—referred to as the "Brain Age Gap", "Brain Age Gap Estimate", or "brain-predicted age difference" (brain-PAD)—serves as an indicator of structural brain health. Studies have shown the brain age paradigm to be sensitive to a wide range of neurological, psychiatric, and metabolic disorders, often revealing a positive Brain Age Gap in conditions like Alzheimer's disease, migraine, depression, and obesity (Beheshti et al., 2020; Han et al., 2021; Navarro-González et al., 2023; Ronan et al., 2016). Conversely, social and lifestyle factors such as higher levels of education, regular physical exercise, playing musical instruments, or engaging in meditation have been associated with a negative Brain Age Gap, suggesting potential protective effects (Luders et al., 2016; Rogenmoser et al., 2018; Steffener et al., 2016).

In schizophrenia, the Brain Age Gap, derived from T1-weighted brain MRI, has consistently been shown to be larger than in healthy individuals (Constantinides et al., 2023; Kim et al., 2023; Koutsouleris et al., 2014; Nenadić et al., 2017; Schnack et al., 2016; Zhu et al., 2023). Common age-related alterations are illustrated in figure 1. However, findings regarding First Episodes (FE) are mixed. Some studies report a significantly higher Brain Age Gap (Hajek et al., 2019; Kim et al., 2023; Mcwhinney et al., 2021), while others find no significant differences (Salisbury et al., 2024).

An interesting result found when analyzing both grey matter (Schnack et al., 2016) and white matter (Wang et al., 2021) is that age mediates the presence of significantly larger Brain Age Gap in schizophrenia patients. In these studies, a larger Brain Age Gap was not found in young patients (Wang et al., 2021). In patients, the Brain Age appears to accelerate around the time of the onset, and its cumulative effect leads to a maximal Brain Age Gap for schizophrenia patients around 5 years later before stabilizing (Schnack et al., 2016). This pattern may help explain why a larger Brain Age Gap is not consistently found in FE.

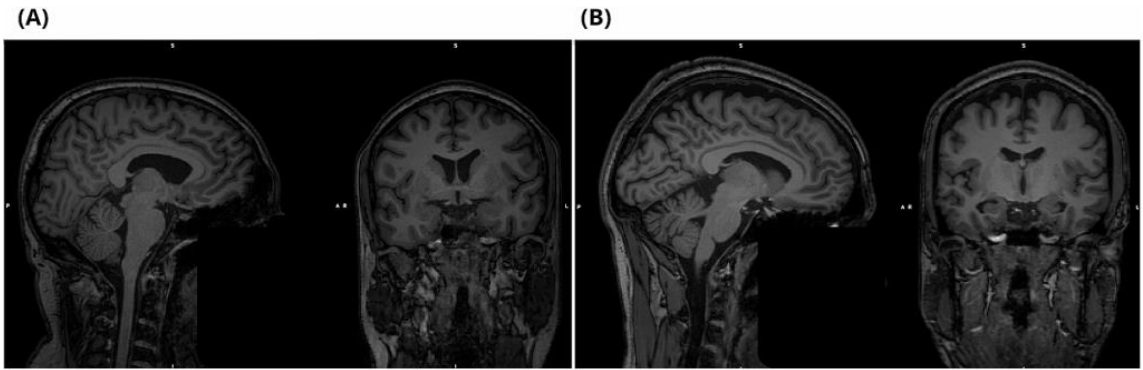

*Fig. 1. Sagittal and coronal MRI of: (A) 42 years old healthy subject, and (B) 44 years old schizophrenia patient. A generalized loss of grey matter can be seen in the central and frontal cortex of the patient. The patient has a 6.6 Brain Age Gap, while the control subject has a -2.6 Brain Age Gap.*

A related but distinct consideration is the effect of antipsychotic medication and patients' lifestyle on the Brain Age Gap. Progressive brain alterations in schizophrenia are also a consequence of antipsychotic medication (Moncrieff and Leo, 2010; Yang et al., 2021), its secondary effects and/or the lifestyle that some patients have, mainly sedentarism and drug abuse (Zipursky et al., 2013). Along these lines, (Mcwhinney et al., 2021) performed a longitudinal study where Brain Age Gap was calculated for first episodes of psychosis at baseline and follow-up. They found that Body Mass Index (BMI) predicted the acceleration rate of Brain Age between scans for both patients and controls. Besides, at follow-up, medication predicted Brain Age Gap once the effect of BMI was controlled. However, the authors propose that medication may not be the cause of a higher Brain Age Gap.

In this article, we aim to sequentially address these issues. First, we aim to replicate the findings in the literature regarding a larger Brain Age Gap in schizophrenia patients. Second, we will try to elucidate whether this Brain Age Gap is present at the beginning of the disease, or whether it depends on its duration. Finally, we will address the possible effect of antipsychotic medication on this Brain Age Gap. The sequence of the research and the statistical methods used can be seen in Figure 2 and in the supplementary material. These issues will be elaborated in the methods section.

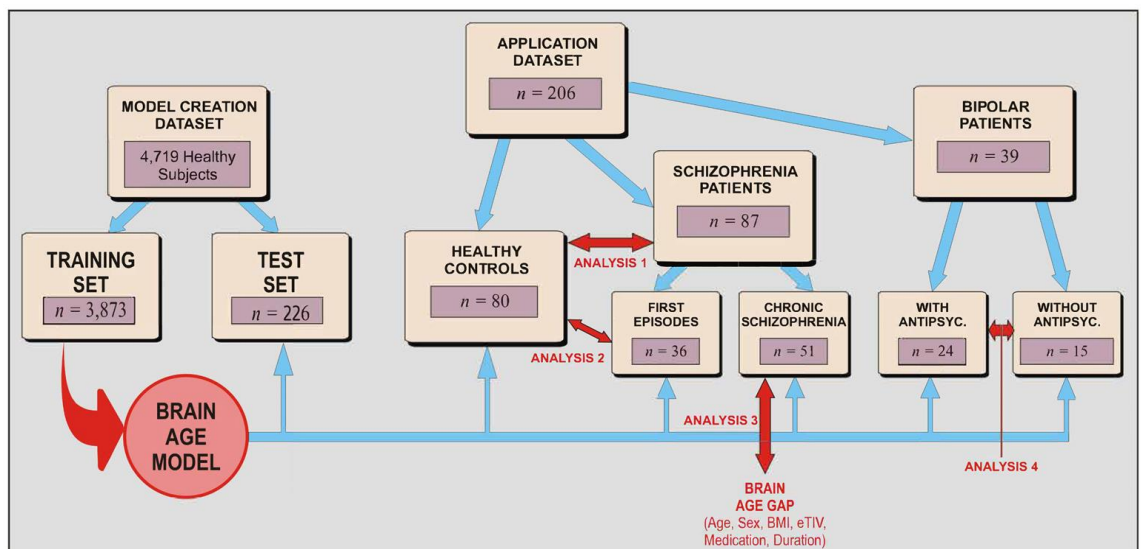

*Fig. 2. Graphical summary of the samples and analyses of this research.*

## 2. Materials and Methods
### 2.1. Brain age model

To create and evaluate our Brain Age model, we compiled a dataset (hereinafter referred to as the *Model Creation Dataset*) consisting of structural T1w MRI scans from 4,065 healthy subjects, drawn from our own database as well as from publicly available studies and repositories. These include: the Dallas Lifespan Brain Study (DLBS) (Park et al., 2024); the Consortium for Reliability and Reproducibility dataset (CoRR) (Zuo et al., 2014); the Neurocognitive aging data release (NeuroCog) (Spreng et al., 2022); The OASIS-1 dataset (Marcus et al., 2007); the Southwest University Adult Lifespan Dataset (SALD) (Wei et al., 2018); the Information eXtraction from Images dataset (IXI) (Biomedical Image Analysis Group, 2023); the CamCAN repository (available at http://www.mrc-cbu.cam.ac.uk/datasets/camcan/) (Shafto et al., 2014; Taylor et al., 2017), the Age Risk dataset (Tisdall and Mata, 2023), the NIMH intramural healthy volunteer dataset (Nugent et al., 2024), the MR_ART dataset (Pardoe and Martin, 2021) and the Nathan Kline Institute (NKI) Rockland Sample. Individuals who had neurological or psychological diagnoses or cognitive impairments were removed from the OASIS, NKI and CoRR datasets. Subjects under 18 and above 65 were also removed to maintain the same age range as our patients and controls sample. The 3,839 images from public databases were used for training and validation, while the 226 images from our database were used as the test set. Supplementary material presents the age and sex distribution of the datasets included in the "Model Creation Dataset".

From the T1w images, we used FastSurfer (Henschel et al., 2020) to extract a total of 1,479 features based on the Desikan-Killiany atlas (Desikan et al., 2006; Klein and Tourville, 2012), from which 175 volumetric features were selected for model input. Volumetric features were prioritized due to their established influence on Brain Age Gap in schizophrenia (Ballester et al., 2023), greater robustness compared to intensity-based features and easier interpretability. We have applied ComBat-GAM to remove scanner effects while preserving biological variation related to age, sex, and eTIV; age was modeled as a smooth term, whereas sex and eTIV were modeled linearly (Pomponio et al., 2020). See Supplementary materials for a deatiled description (Pomponio et al., 2020).

We selected a *transformer*-based regressor model, originally developed by Vaswani et al. (Vaswani et al., 2017) and here adapted to our specific scenario. Due to its powerful attention mechanism, *transformer*-based architectures enhance the ability to capture long-distance relationships within the input data. The original architecture includes an encoder-decoder framework originally designed for natural language processing. However, for this regression task we have employed only the encoder to model the relationships within the data, and the encoder's output was then fed into a regression model.

We implemented the model in PyTorch (Paszke et al., 2019) using the Anaconda 2.4.1 environment. The model implements a TabTransformer encoder tailored for tabular regression over 175 numerical features. Each feature is first mapped independently into a 32-dimensional token via a small feed-forward block (Linear (1→32) + ReLU + Dropout (0.1)), yielding 175 tokens of size 32. These tokens are then processed by two stacked encoder modules: each layer performs multi-head self-attention over the 175 token sequence, followed by a two-stage feed-forward network with residual connections, layer normalization, and dropout to stabilize and regularize training. After the encoder, the output tokens are concatenated into a single 5600-dimensional vector (175×32) and fed through a final Linear regression head to predict the continuous target (e.g., age). The model was optimized with Adam and the MSELoss function and trained for 500 epochs. In order to measure and report the performance of the Brain Age model, we used the mean absolute error (MAE) and the Pearson's correlation coefficient ($r$). Metrics will be also presented separately for males and females to ensure the model does not exhibit sex bias. To enhance the robustness of our findings we replicated the statistical analyses with the brain age derived from Pyment model (Leonardsen et al., 2022). The Pyment model is an open-source deep learning model which uses the entire MRI image as input. Since the two models are based on fundamentally different approaches their predictions are likely driven by distinct sets of features or patterns. Therefore, any convergence in their results would provide stronger and more robust evidence for the observed effects.

### 2.2. Participants

A total of 87 schizophrenia patients, 39 bipolar disorder patients and 80 controls were included in this study. This dataset will be hereinafter referred to as the *Application Dataset*. A summary of controls and schizophrenia patients can be seen in Table 1, while a summary of bipolar patients is described in Table 2. A detailed description of all patients' psychopharmacological prescriptions can be found in the Supplementary Material.

Exclusion criteria were: (a) intelligence quotient below 70, (b) present or past substance dependence (excluding caffeine and nicotine), (c) head trauma with loss of consciousness, (d) neurological or mental diagnosis other than schizophrenia, and (e) any other treatment affecting the central nervous system. All participants provided written informed consent after receiving comprehensive written information. The local ethics committee approved the study. This work complies with the ethical standards of the Helsinki Declaration of 1975, as revised in 2008.

*Table 1 Demographic and clinical characteristics of controls and schizophrenia patients. These subjects were included in (controls and patients), analysis 2 (controls and first episodes) and analysis 3 (chronic patients).*

|  | **Controls** | **Patients** | **First Episodes** | **Chronic Schizophrenia** |
|---|---|---|---|---|
| **Sample size** | 80 | 87 | 36 | 51 |
| **Age, years** | 31.58 (11.93) | 35.24 (11.04) * | 29.47 (9.48) | 39.31 (10.29) |
| **Sex, M/F** | 44/36 | 49/38 | 19/17 | 30/21 |
| **Body Mass Index** | 24.19 (4.1) | 25.5 (5.4) | 22.91 (3.5) | 27.32 (5.8) |
| **Estimated total intracranial volume, liters** | 1.57 (1.50) | 1.54 (1.71) | 1.55 (1.80) | 1.53 (1.65) |
| **Illness duration, months** | – | 80.74 (118.40) | 11.34 (18.35) | 133.54 (134.43) |
| **Diagnoses S/FE** | – | 51/36 | 326 (200.49) | 435.14 (271.12) |
| **CPZ equivalents** | – | 390 (249.03) |  |  |

* Significant difference with *p*-value < 0.05.

*Table 2 Demographic and clinical characteristics of the bipolar patients for analysis 4.*

|  | **Bipolar with Antipsychotics** | **Bipolar without Antipsychotics** | **Test statistic** | ***p*-value** |
|---|---|---|---|---|
| **Sample size** | 24 | 15 | - | - |
| **Age, years** | 39.29 (11.72) | 47.49 (11.48) | $t = 1.02$ | 0.31 |
| **Sex, M/F** | 13/11 | 8/7 | $\chi^2 = 0.03$ | 0.98 |
| **Body Mass Index** | 30 (6) | 27.24 (5.5) | $t = 1.7$ | 0.09 |
| **Estimated total intracranial volume, liters** | 1.56 (1.30) | 1.50 (1.59) | $t = 0.56$ | 0.58 |
| **Illness duration, months** | 179.09 (117.72) | 231.36 (140.11) | – | – |
| **CPZ equivalents** | 291.67 (144.66) | – | – | – |

### 2.3. MRI acquisition

For the *Application Dataset*, high-resolution 3D T1-weighted MRI data were acquired using a Philips Achieva 3T MRI unit (Philips Healthcare, Best, The Netherlands) with a 32-channel head. For the anatomical T1-weighted images, acquisition parameters were: Turbo Field Echo (TFE) sequence, repetition time (TR) = 8.1 ms, echo time (TE) = 3.7 ms, flip angle = 8°, 256 × 256 matrix size, 1 × 1 × 1 mm3 spatial resolution and 160 slices covering the whole brain. Following the image acquisition, image segmentation and feature extraction were performed as described for the *Model Creation Dataset*.

### 2.4. Statistical analysis

Brain Age models, as other regression models, usually suffer from regression to the mean, a tendency to overestimate the predicted age for younger individuals and underestimate the predicted age for older individuals. Thus, predicted age in our model was corrected using the procedure proposed by (Cole et al., 2018). The Pyment model has no built-in regression to the mean correction and original training data is needed to perform (Cole et al., 2018) correction. Therefore, predicted age for the Pyment model was not corrected. For both cases age has been included as covariable into the analysis to remove further age bias of the models. These analyses were:

Analysis 1, schizophrenia vs controls: In order to replicate the findings in the literature of a larger Brain Age Gap in schizophrenia patients, we used ANCOVA with Brain Age Gap as dependent variable, disease condition as grouping variable and age, sex, BMI and eTIV as covariables.

Analysis 2, FE vs controls: In order to compare healthy controls against FE, we used an ANCOVA with the same covariables as in Analysis 1.

Analysis 3, predictors of Brain Age Gap: In order to assess the impact of duration and medication on the Brain Age Gap we used a multivariate regression analysis with Brain

Age Gap as dependent variable. Age, sex, BMI, eTIV, disease duration, chlorpromazine equivalents and interaction between duration and medication were included as covariates. For this analysis, only chronic patients were employed. As we are lacking complete follow-up records of medication use, we included three proxy variables of cumulative exposure in a multiple regression model: (1) current antipsychotic dosage (chlorpromazine equivalents), (2) the interaction between dosage and illness duration, and (3) BMI, given its link to weight gain as a side effect.

Analyses 1, 2 and 3 are based on linear models, and therefore their assumptions need to be fulfilled (Leppink, 2018). All assumptions are met except for the homoscedasticity of the variance for analyses 1 and 3. Therefore, we resorted to the Halbert-White (White, 1980) method, since it is not sensitive to heteroscedasticity. Analysis 1 assumption of normality of the residuals was not met. A visual examination of the Q-Q plot proved that it was due to an outlier among the schizophrenia patients. Normality of residuals was assessed after excluding this outlier from the analysis. Besides, results were virtually the same with or without this outlier, with only negligible changes in the coefficients. All the information on statistical assumptions can be found in the Supplementary Material. The results presented in the manuscript, based on our transformer model, exclude this outlier.

Analysis 4, effects of current antipsychotic treatment in bipolar disorder: In order to further assess the influence of medication on the Brain Age Gap we compared bipolar patients with and without current antipsychotic treatment. The sample size limited the statistical analysis that could be applied, besides, the residuals of the linear model showed a non-normal behavior (Shapiro-Wilk's p-value = 0.05). Therefore, a Mann-Whitney's U test was be used for this comparison.

Variables were normalized before all analyses which were implemented using Statsmodels (Seabold and Perktold, 2010). The dataset containing the main data supporting the present results is available (Mendeley Data DOI: 10.17632/nz2wnz3vhk.1). The script containing the main analyses are available on https://github.com/alerohe/brain_age_schizophrenia.

## 3. Results

We first assessed the accuracy of our Brain Age model. On the test set of the *Model Creation Dataset*, our model obtained MAE: 6.41 Pearson's $r$: 0.76. The performance metrics by sexes were. MAE=5.30, Pearson $r$=0.80 for males, and MAE=7.20, $r$=0.83 for females. Performance was also good on healthy controls of the Application Dataset (MAE = 5.42 and a $r$ = 0.83). After the correction of the regression towards the mean issue we obtained MAE = 5.47 and a $r$ = 0.83. Pyment model's performance was superior in the Application Dataset's controls (MAE = 3.91, $r$ = 0.93). Graphical results can be seen in Figure 3.

Analysis 1, schizophrenia vs controls: a) Transformer-based model: The ANCOVA's size effect was moderate-large, with Cohen's $f^2$ = 0.29. Brain Age Gap was significantly higher in patients, $z$ = 4.70, $p$-value = 0.000004. The Brain Age gap mean difference between patients and controls is 5.8 years. eTIV also significantly influenced the Brain Age Gap with $z$ = -2.49, $p$-value = 0.01. b) Pyment model: The ANCOVA's size effect was large, with Cohen's $f^2$ = 0.43. Brain Age Gap was significantly higher in patients, $z$

= 4.36, *p*-value = 0.00001. The Brain Age gap mean difference between patients and controls is 5.92 years. Age also significantly influenced the Brain Age Gap with *z* = -6,3 *p*-value = 0.0000000002.

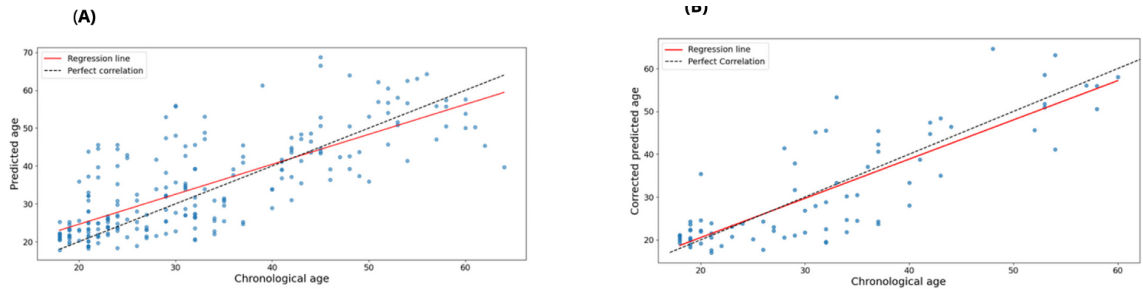

*Fig. 3. Perfect correlation lines and regression lines between predicted age and chronological age for: (A) test set of the Model Creation Dataset, and (B) Application dataset's controls with regression towards the mean correction.*

Analysis 2, FE vs controls comparison: a) Transformer-based model: The ANCOVA's size effect was moderate, Cohen's f² = 0.17. Brain Age Gap was significantly different between FE patients and controls, *z* = 3.076, *p*-value = 0.002. The Brain Age gap mean difference between FE patients and controls is 4.34 years. eTIV also significantly influenced the Brain Age Gap with *z* = -2.493 *p*-value = 0.013; b) Pyment model: The ANCOVA's size effect was large, with Cohen's f² = 0.63. Brain Age Gap was significantly higher in patients, *z* = 3.95, *p*-value = 0.00007. The Brain Age gap mean difference between FE patients and controls is 9.93 years. Age also significantly influenced the Brain Age Gap with *z* = -3,90, *p*-value = 0.0001.

Analysis 3, predictors of Brain Age Gap: a) Transformer-based model: As seen in Table 3, none of the predictors were significantly associated with the Brain Age Gap. Coefficients of chlorpromazine equivalents and duration were low, indicating no meaningful relationship. b) Pyment model: Age was the only significant regressor. Coefficients of chlorpromazine equivalents and duration were low, indicating no meaningful relationship, while its interaction showed a small positive correlation with the Brain Age Gap.

*Table 3 Summary of the regression model used in Analysis 3 based on age predictions from our transformer-based model.*

| Predictor | Beta Coefficient | Standard Error | z | *p*-value |
|---|---|---|---|---|
| Intercept | 0.63 | 0.37 | 1.73 | 0.08 |
| Chlorpromazine equivalents | –0.01 | 0.24 | –0.07 | 0.94 |
| Duration | 0.12 | 0.33 | 0.7 | 0.83 |
| Chlorpromazine Equivalents × Duration | –0.05 | 0.26 | –0.21 | 0.83 |
| Age | 0.28 | 0.24 | –1.18 | 0.23 |
| Sex | –0.37 | 0.41 | –0.90 | 0.37 |
| eTIV | –0.39 | 0.23 | –1.68 | 0.09 |
| BMI | 0.23 | 0.21 | 1.13 | 0.25 |

*Standard errors are heteroscedasticity-consistent (HC3). Model fit: $R^2 = 0.23$, $F(7, 37) = 1.67$, $p = 0.15$.*

*Table 4 Summary of the regression model employed in Analysis 3 using Pyment's age prediction.*

| Predictor | Beta Coefficient | Standard Error | z | *p*-value |
|---|---|---|---|---|
| Intercept | 0.58 | 0.39 | 1.45 | 0.15 |
| Chlorpromazine Equivalents | –0.04 | 0.28 | –0.15 | 0.88 |
| Duration | –0.05 | 0.33 | –0.17 | 0.86 |
| Chlorpromazine Equivalents × Duration | 0.18 | 0.51 | 0.37 | 0.71 |
| Age | –1.12 | 0.76 | –2,98 | 0.003 |
| Sex | –0.47 | 0.55 | –0.86 | 0.40 |
| eTIV | –0.04 | 0.26 | –0.15 | 0.88 |
| BMI | 0.21 | 0.15 | 1.47 | 0.14 |

*Standard errors are heteroscedasticity-consistent (HC3). Model fit: $R^2 = 0.47$, $F(7, 38) = 5.38$, $p = 0.0002$.*

Analysis 4, effects of current antipsychotics in bipolar disorder: a) Transformer-based model: The Mann-Whitney's U test effect size was small, $r = 0.06$. The U-stat was 185 with a *p*-value of 0.89. Bipolar patients under antipsychotic treatment showed a similar Brain Age Gap, with a mean difference of approximately 0.3 years; b) Pyment model: The Mann-Whitney's U test effect size was small, $r = 0.12$. The U-stat was 206 with a *p*-value of 0.46. Bipolar patients did not show any significant difference, regardless of antipsychotic medication. The mean difference of approximately 6 years. Note that Mann Whitneys' U test is not a mean-based statistic, therefore, although the mean differences are similar for both models' prediction, the group comparisons may yield different results. Boxplots for controls, all schizophrenia patients, FE and bipolar patients can be seen in Figure 4.

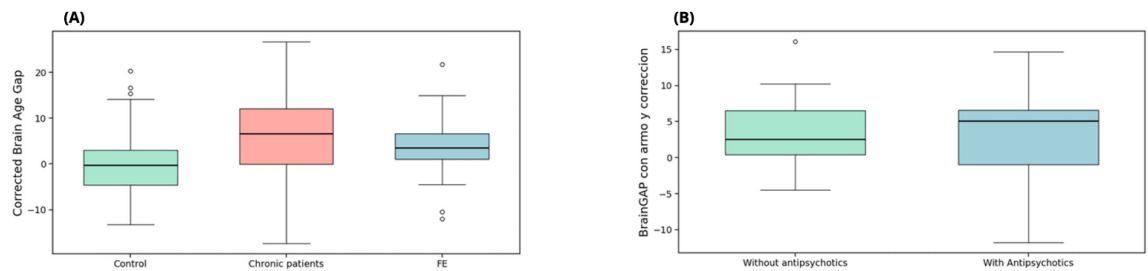

*Fig. 4. Boxplots for: (A) controls, all schizophrenia patients and FE, and (B) bipolar disorder patients.*

## 4. Discussion

Our first analysis confirmed the presence of a larger Brain Age Gap in schizophrenia patients compared to healthy controls. As noted in the introduction, this result is consistent with previous findings. Contrary to prior reports, BMI did not emerge as a significant covariate in the ANCOVA, contributing to an increased Brain Age Gap in both groups (Chin Fatt et al., 2021). For the Pyment model the results also showed higher Brain Age Gap in patients while age was a significant covariate. The lack of regression to the mean correction is likely responsible for the consistent age effect observed across analyses.

Our second analysis investigated whether an increased Brain Age Gap is already present at the onset of the illness. Our model and Pyment found larger Brain Age Gap in FE patients. Previous literature report both a larger Brain Age Gap (Hajek et al., 2019; Kim et al., 2023; Mcwhinney et al., 2021; Yi-Bin et al., 2022) and no difference compared to controls (Salisbury et al., 2024). There could be several factors contributing to this inconsistency in the literature. According to a recent meta-analysis, when present, differences between FE and controls are smaller than those observed in chronic patients (Blake et al., 2023). In addition to the wide variety of machine learning models and input features, these smaller differences may be more sensitive to methodological variations than the comparisons between chronic patients and healthy controls.

Regarding FE patients, it is also important to highlight that previous longitudinal research has shown that the Brain Age Gap emerges after disease onset and tends to increase during the first five years following illness onset (Schnack et al., 2016). Interestingly, however, disease duration did not predict the Brain Age Gap in chronic schizophrenia patients (analysis 3), which could suggest a nonlinear relationship between Brain Age Gap and disease duration, as also suggested by (Schnack et al., 2016).

To investigate this nonlinear relationship, we have replicated the analyses allowing a nonlinear fitting of the covariables and regressors. The main results are roughly equal to the linear analyses, although a significant nonlinear relationship between BMI and Brain Age Gap was found when introduced as a covariable for group comparisons (analyses 1 and 2). A full description of these results is provided in the supplementary material.

Although not included in the methodology of the present study, it is of great importance to address the heterogeneity of schizophrenia (Tandon et al., 2024). Previous work has shown that schizophrenia patients exhibit greater variability in the volumes of the temporal cortex, thalamus, putamen, and third ventricle compared to controls (Brugger and Howes, 2017). According to our results, this is also true for the brain age framework. The whiskers of the boxplots in Figure 4 are broader for patients, especially in the Pyment model. When focused on structural brain characteristics, recent work on international datasets has identified two separate clusters within schizophrenia (Jiang et al., 2024a, 2023), and different epicenters of disease origin (Jiang et al., 2024b). One subgroup has an early generalized cortical deficit, especially in the Broca's Area, while the other subgroup has an early subcortical deficit. Similarly, when the clustering variables comprise the cognitive domain, one of the subgroups also presents a larger structural atrophy in subcortical regions (Fernández-Linsenbarth et al., 2021). Therefore, subtypes of schizophrenia present distinct patterns of structural brain atrophy, primarily at the beginning of the disease. These anatomical subtypes may influence the Brain Age Gap and help explain some of the inconsistencies observed in FE previous literature. Brain age research in schizophrenia subtyping remains scarce (Haas et al., 2022) and further research is needed to draw conclusions.

Additionally, it is possible that some of these divergent patterns reflect not only pathological changes, but also compensatory mechanisms (Palaniyappan, 2023). Recent work has proposed that certain structural alterations in schizophrenia—such as region-specific cortical thickening or shifts in network centrality—may represent adaptive responses to the illness, rather than direct markers of damage (Guo et al., 2016;

Palaniyappan et al., 2019). Recognizing this possibility could help explain some of the variability in Brain Age estimates across patients.

Another relevant feature of analyses 1 and 2 is the significant influence of eTIV on the Brain Age Gap. We conducted the same analysis excluding eTIV from the ANCOVA. Afterwards, sex became significant in all analyses. Our model's predictions for males and females differed by 1.9 years in the test set. Taken together, this pattern suggests that part of the eTIV–Brain Age Gap association may reflect sex-related differences in head size and model behavior, given that males have larger eTIV on average (Ruigrok et al., 2014). Besides, eTIV is not relevant in the Pyment model (see coefficients in analysis 3), which shows no sex bias. We consider that the inclusion of eTIV and sex as covariates validates the findings in our model, further supported by the similar results obtained using Pyment. Nonetheless, it provides evidence of the relevance of sex differences in the Brain Age framework.

Our third and fourth analyses aimed to clarify the potential effects of antipsychotic treatment on the Brain Age Gap. Antipsychotics are known to affect brain structure, including reductions in cortical thickness—particularly in the parietal lobe—and increases in basal ganglia volume (Emsley et al., 2023; Huhtaniska et al., 2017). However, estimating cumulative exposure is challenging, as treatment adherence rates remain below 70% according to electronic monitoring studies (Yaegashi et al., 2020). To address this, we included three proxy variables for cumulative exposure in a multiple regression model: (1) current antipsychotic dosage (chlorpromazine equivalents), (2) the interaction between dosage and illness duration, and (3) BMI, given its link to weight gain as a side effect. None of these variables was significantly associated with Brain Age Gap—neither in our transformer-based model, nor in Pyment. Independently of the model, BMI showed a positive but non-significant association with the Brain Age Gap. Since BMI has been previously associated with increased Brain Age Gap in healthy individuals, it is possible that the impact of antipsychotics on brain aging is mediated by their obesogenic effects (Mcwhinney et al., 2021). In the post-hoc non-linear replication of this analysis, neither age nor any other regressors were significant.

To provide additional evidence regarding the effects of antipsychotic medication, we also conducted a Mann-Whitney U test comparing bipolar patients receiving antipsychotic treatment with those not receiving it. The absence of a significant difference in both models supports the view that antipsychotics have no effect on Brain Age Gap. To further strengthen the evidence for the null hypothesis in this analysis we have conducted a Bayesian Mann-Whitney test, which yielded similar results, —see Supplementary Materials. Although indirect, our proxies for medication use and accumulation provide compelling evidence supporting either a null effect or even a potentially protective role of antipsychotic medication in brain aging.

This study has several limitations. First, its cross-sectional design limits the ability to draw causal conclusions, particularly regarding age-related changes. A longitudinal design with a placebo-controlled group would be more appropriate to assess the specific effects of antipsychotic medication on brain aging. Second, our proxies for medication exposure do not capture the pharmacological differences between antipsychotics and may overlook variations in treatment adherence and antipsychotic class. Third, the sample size of bipolar patients was limited, and replication in larger cohorts is necessary to validate the observed protective effect. Fourth, although lifestyle was partially controlled for through BMI, other relevant factors—such as smoking or physical activity, both known to influence brain aging (Bittner et al., 2021), were not included. Fifth, although it requires larger sample sizes, training sex-specific models is less biased than using a unified one. Finally, future studies should consider differences in treatment response, given the reported relation between brain age and antipsychotic response in this group (Fan et al., 2025).

**Acknowledgements**

This work was supported by research grants PRE2022-104038 funded by MCIN/AEI/10.13039/501100011033 and the ESF+, and PID2021-124407NB-I00, funded by MCIN/AEI/10.13039/501100011033/FEDER, UE.

The authors are profoundly thankful to the owners of the Neurocognitive aging data release, which is available at OpenNeuro under the identifier "OpenNeuro Dataset ds003592.", the IXI dataset available at https:// brain-development. org/ ixi-datas et/, the DLBS dataset and the International Neuroimaging Data-Sharing Initiative Group and the participants of the Consortium for Reliability and Reproducibility (CoRR) for sharing their data publicly so this research was made possible.

Data were provided in part by OASIS OASIS-1: Cross-Sectional: Principal Investigators: D. Marcus, R, Buckner, J, Csernansky J. Morris; P50 AG05681, P01 AG03991, P01 AG026276, R01 AG021910, P20 MH071616, U24 RR021382; Data collection and sharing in part for this project was provided by the Cambridge Centre for Ageing and Neuroscience (CamCAN). CamCAN funding was provided by the UK Biotechnology and Biological Sciences Research Council (grant number BB/H008217/1), together with support from the UK Medical Research Council and University of Cambridge, UK. The SALD data repository was supported by the National Natural Science Foundation of China (31470981; 31571137; 31500885), National Outstanding young people plan, the Program for the Top Young Talents by Chongqing, the Fundamental Research Funds for the Central Universities (SWU1509383, SWU1509451, SWU1609177), Natural Science Foundation of Chongqing (cstc2015j-cyjA10106), Fok Ying Tung Education Foundation (151023), General Financial Grant from the China Postdoctoral Science Foundation (2015M572423, 2015M580767), Special Funds from the Chongqing Postdoctoral Science Foundation (Xm2015037, Xm2016044), Key research for Humanities and social sciences of Ministry of Education (14JJD880009).

We also acknowledge the Child Mind Institute for access to the Nathan Kline Institute (NKI) Rockland Sample, the National Institute of Mental Health of the USA for the access to the intramural healthy volunteer dataset and the Brain Imaging Centre of Budapest for the access to the MR_ART dataset.

In addition, we are extremely grateful to all the patients and participants who volunteered to take part in this study.

**Contributors**

A. Roig and L.M. San-José wrote the draft and performed the statistical analysis; V. Molina, recruited patients; L.M. San-José and R. Navarro developed the Brain Age Model; R. de Luis, V. Molina designed the study; R. de Luis, V. Molina and L.M. San-José revised the draft.

**Appendix A. Supplementary data**

Supplementary data to this article can be found online at https://doi.org/10.1016/j.schres.2025.11.008.